\documentclass[pra,twocolumn,floatfix,showpacs]{revtex4-1}
\usepackage{graphicx,color}
\usepackage{amsmath,amssymb,bm}

\newcommand{\tn}{\textnormal}
\newcommand{\cpra}[3]{Phys.~Rev.~A {\bf #1}, #2 (#3)}
\newcommand{\cprb}[3]{Phys.~Rev.~B {\bf #1}, #2 (#3)}
\newcommand{\cprl}[3]{Phys.~Rev.~Lett.~{\bf #1}, #2 (#3)}
\newcommand{\cnjp}[3]{New J.~Phys.~{\bf #1}, #2 (#3)}

\newcommand{\cbook}[2]{\textit{#1} (#2)}
\newcommand{\cjpa}[3]{J.~Phys.~A {\bf #1}, #2 (#3)}
\newcommand{\cepj}[3]{Eur.~Phys.~J.~Spec.~Topics {\bf #1}, #2 (#3)}
\newcommand{\cjpamg}[3]{J.~Phys.~A: Math.~Gen {\bf #1}, #2 (#3)}
\newcommand{\cprd}[3]{Phys.~Rev.~D {\bf #1}, #2 (#3)}
\newcommand{\cpre}[3]{Phys.~Rev.~E {\bf #1}, #2 (#3)}

\definecolor{darkred}{rgb}{0.90,0,0}
\definecolor{darkgreen}{rgb}{0,0.60,.2}
\definecolor{darkblue}{rgb}{0,0,1}
\definecolor{grey}{cmyk}{0,0,0,0.25}
\definecolor{orange}{cmyk}{0,0.6,0.8,0}

\begin{document}

\title{Non-equilibrium thermal transport and vacuum expansion in the Hubbard model}

\author{C.\ Karrasch}

\affiliation{$^2$Dahlem Center for Complex Quantum Systems and Fachbereich Physik, Freie Universit\"at Berlin, 14195 Berlin, Germany}

\begin{abstract}

One of the most straightforward ways to study thermal properties beyond linear response is to monitor the relaxation of an arbitrarily large left-right temperature gradient $T_L-T_R$. In one-dimensional systems which support ballistic thermal transport, the local energy currents $\langle j(t)\rangle$  acquire a non-zero value at long times, and it was recently investigated whether or not this steady state fulfills a simple additive relation $\langle j(t\to\infty)\rangle=f(T_L)-f(T_R)$ in integrable models. In this paper, we probe the non-equilibrium dynamics of the Hubbard chain using density matrix renormalization group (DMRG) numerics. We show that the above form provides an effective description of thermal transport in this model; violations are below the finite-time accuracy of the DMRG. As a second setup, we study how an initially equilibrated system radiates into different non-thermal states (such as the vacuum).

\end{abstract}

\maketitle



\section{Introduction}

Low-dimensional electron systems govern the behavior of strongly anisotropic materials (such as the iron pnictides), of graphene, or of quantum dots and wires. 1d or 2d fermions or bosons can also be realized in cold atom setups where elementary equilibrium or real time physics can be probed accurately. Thus, understanding low-dimensional systems is important both fundamentally and for applications in nanoelectronics or the design of functional materials. However, Coulomb interactions lead to a variety of many-body phenomena that cannot be obtained by using simple perturbation theory but require more elaborate analytical or numerical techniques. It is particularly challenging to treat systems which are not in thermal equilibrium.

One area that has attracted considerable attention during the past decades is transport in one dimension where interacting models exist that can be diagonalized exactly using Bethe ansatz techniques; typical examples are XXZ spin chains or the Fermi-Hubbard model \cite{giamarchi,hubbook}. In these systems, local conserved charges can effectively prevent currents from scattering and can thus in principle lead to dissipationless transport at finite temperature $T>0$ \cite{andrei,sirker,integrability}. However, even if the whole spectrum of the Hamiltonian is known, it remains a formidable task to determine the linear response conductivity quantitatively since it is governed by the couplings between all excited states \cite{bethespec2,bethespec}, which are difficult to compute in the Bethe ansatz \cite{bethespin}. Significant progress has been made within recent years using novel analytical \cite{prosen1,prosen2} or numerical \cite{drudepaper,steinigeweg,steinigeweg2} techniques.

\begin{figure}[b]
\includegraphics[width=0.8\linewidth,clip]{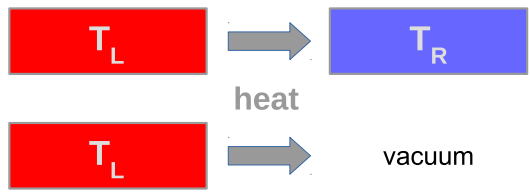}
\caption{(Color online) The main setups considered in this paper. A Fermi-Hubbard chain is prepared in thermal equilibrium at a temperature $T_L$. At time $t=0$, it is connected to a vacuum state or to a chain exhibiting a different temperature $T_R$. The former describes the expansion after switching off a (sharp) potential barrier, the latter models the evolution of an initial temperature gradient.}
\label{fsetup}
\end{figure}

Studying transport out of equilibrium is complicated and constitutes one of the most active areas of research in strongly correlated condensed matter physics (for a non-comprehensive list of recent works in this direction, see \cite{noneqtherm1,noneqtherm2,schmitteckert,fabiannoneq,fabiannoneq2,noneqprosen1,noneqstein,znidaric1,znidaric2,znidaric3,prosentherm,noneqjaksch,finiteTquench,typicalitynoneq,mendl,fagotti2,field1}). A simple way to probe 1d thermal transport \cite{fabianprb,fabianrev,thermconserved,bethetherm,qmctherm,roschtherm,orignac,saito,chernyshyev,rosch3,energyhubbard} beyond the regime of linear response is to monitor the real time evolution of an initial, arbitrarily large temperature gradient. Despite its general simplicity and experimental realization in quasi-1d spin systems \cite{ott,ott2,hess,hess2}, this setup had not been studied theoretically until a few years ago \cite{free2,doyon,thermalpaper,bruneau} but has gained considerable attention since then \cite{doyon7,prelovsek,doyon3,eisler1,martelloni,romainpaper,jaksch,rossini,doyon2,viti1,doyon5,rossini2,fagotti,doyon4,field2,doyon6,zotos}. The aim of this paper is to extend previous works to the 1d Fermi-Hubbard model and to discuss non-equilibrium setups which are more general than temperature gradients (and which might thus be of higher relevance for cold atom experiments).

From a theoretical perspective, the simplest way to create a $T$-gradient is to prepare two semi-infinite `left and right' chains in equilibrium at temperatures $T_{L,R}$ and to connect them at time $t=0$. In general, one expects the system to thermalize after some transient time and hence no energy current $\langle j(t)\rangle$ to flow across the interface for $t\to\infty$. In an integrable model, however, it is reasonable to assume that the conserved charges inhibit scattering for arbitrarily large $T_L-T_R$ and that $\langle j(t)\rangle$ acquires a finite steady-state value. More interestingly, it was first proposed in Ref.~\cite{doyon} in the context of conformal field theory that the asymptotic current is not a complicated function of $T_L$ and $T_R$ but rather has a simple form,
\begin{equation}\label{ffunc}
\lim_{t\to\infty}\langle j(t)\rangle = f(T_L)-f(T_R)\,,
\end{equation}
where $f(T)$ is a model-dependent function of the temperature. The derivative of $f$ is nothing but the \textit{linear reponse} conductance -- which thus determines the current even far \textit{out of equilibrium}. This `additivity property' was confirmed analytically for free lattice fermions \cite{free2,bruneau} and tested using density matrix renormalization group (DMRG) numerics for the XXZ spin chain \cite{thermalpaper}. Its validity for the interacting XXZ chain was subsequently discussed critically in various works \cite{doyon3,rossini,fagotti,zotos,doyon6}; in particular, Eq.~(\ref{ffunc}) is not compatible with the hydrodynamic approaches of Refs.~\cite{fagotti,doyon6}. This suggests that Eq.~(\ref{ffunc}) provides an effective description of transport in the XXZ chain but is violated on a scale that cannot be resolved by the DMRG. It is the first goal of this work to shed more light on these issues by revisiting the temperature-gradient setup for the Fermi-Hubbard model (FHM).

Intuitively, one can view Eq.~(\ref{ffunc}) as a form of `effective thermal radiation' \cite{sotiriadis,cardyprize}. This interpretation suggests that the function $f(T)$ could not only describe $T$-gradients but approximately govern the thermal currents which flow out of an equilibrated `left' chain into \textit{any} state on the right as long as the time evolution is dictated by an integrable Hamiltonian. It is the second aim of this paper to collect evidence whether this hypothesis is true or false. Most importantly, we will study the expansion (into the vacuum) after switching off a sharp potential trap.

This exposition is organized as follows. In Sec.~\ref{sec:model}, we introduce the model, discuss several ways to prepare the initial state, and briefly sketch the idea of density matrix renormalization group calculations. The evolution of temperature gradients in the Hubbard model is studied in Sec.~\ref{sec:T}. In Sec.~\ref{sec:vac}, we show results for the expansion into the vacuum. The XXZ chain is less computationally demanding than the FHM, and our hypothesis can be tested for a larger class of states; we show data in the appendix.

\section{Model and Method}
\label{sec:model}

\subsection{Model}

The one-dimensional Fermi-Hubbard model is governed by $H=\sum_{n}h_n$ with local terms 
\begin{equation}\begin{split}\label{hub}
h_n = &-\frac{t_0}{2} \left(c_{n\uparrow}^\dagger c_{n+1\uparrow}^{\phantom{\dagger}} +c_{n\downarrow}^\dagger c_{n+1\downarrow}^{\phantom{\dagger}} + \tn{h.c.} \right)\\ 
&+ \frac{U}{2}\left(\tilde n_{l\uparrow} \tilde n_{l\downarrow}+ \tilde n_{l+1\uparrow} \tilde n_{l+1\downarrow}\right)\\
&+ V (\tilde n_{l\uparrow}+\tilde n_{l\downarrow}) (\tilde n_{l+1\uparrow}+\tilde n_{l+1\downarrow}) \,,
\end{split}\end{equation}
where $c_{n\sigma}$ annihilates a fermion with spin $\sigma$ on site $n$, and $\tilde n_{n\sigma}=c_{n\sigma}^\dagger c_{n\sigma}^{\phantom{\dagger}}-1/2$. The on-site and nearest-neighbor interactions are denoted by $U$ and $V$, respectively; the system is integrable for $V=0$ \cite{hubbook}. Throughout this paper, we use the hopping matrix element $t_0=1$ as the unit of energy. The local energy current is defined via a continuity equation and reads $j_n=i [h_{n+1},h_{n}]$.

\subsection{State preparation}

The straightforward choice of the initial density matrix $\rho_0$ is given by the product state
\begin{equation}\label{state1}
 \rho_0 \sim \rho_L\otimes\rho_R\,,
\end{equation}
where $\rho_{L,R}$ are the statistical operators governing the isolated left and right systems, respectively. In order to model an initial temperature gradient \cite{free2,bruneau,thermalpaper,doyon}, they are both chosen as thermal density matrices featuring different $T_{L,R}$:
\begin{equation}
\rho_{L} \sim e^{-H/T_{L}}\,,~~\rho_{R} \sim e^{-H/T_{R}}\,.
\end{equation}
Alternatively, one can employ $\rho_R\sim\rho_\tn{vac}$ or $\rho_R\sim e^{-K/T_R}$ in order to study the expansion into the vacuum or into a thermal state of a different Hamiltonian $K$, respectively.

\begin{figure}[t]
\includegraphics[width=0.95\linewidth,clip]{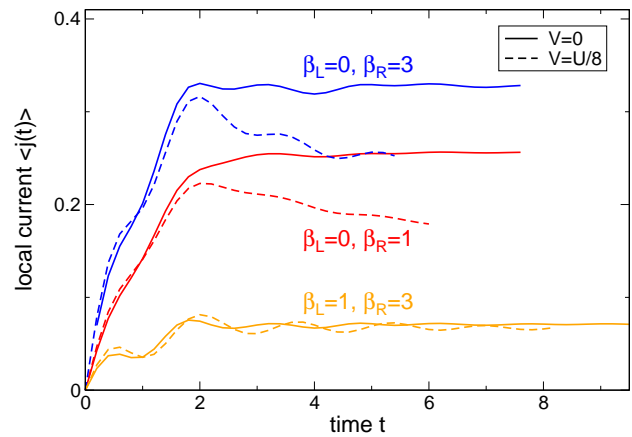}
\caption{(Color online) Thermal current flowing through the Hubbard chain in the presence of an initial temperature gradient defined by $\beta_{L,R}=1/T_{L,R}$. The on-site and nearest-neighbor interaction strength are $U=4$ and $V\in\{0,U/8\}$, respectively. In the integrable case $V=0$, the currents saturate to a non-zero stationary-state value.  }
\label{fcurrent}
\end{figure}

\begin{figure*}[t]
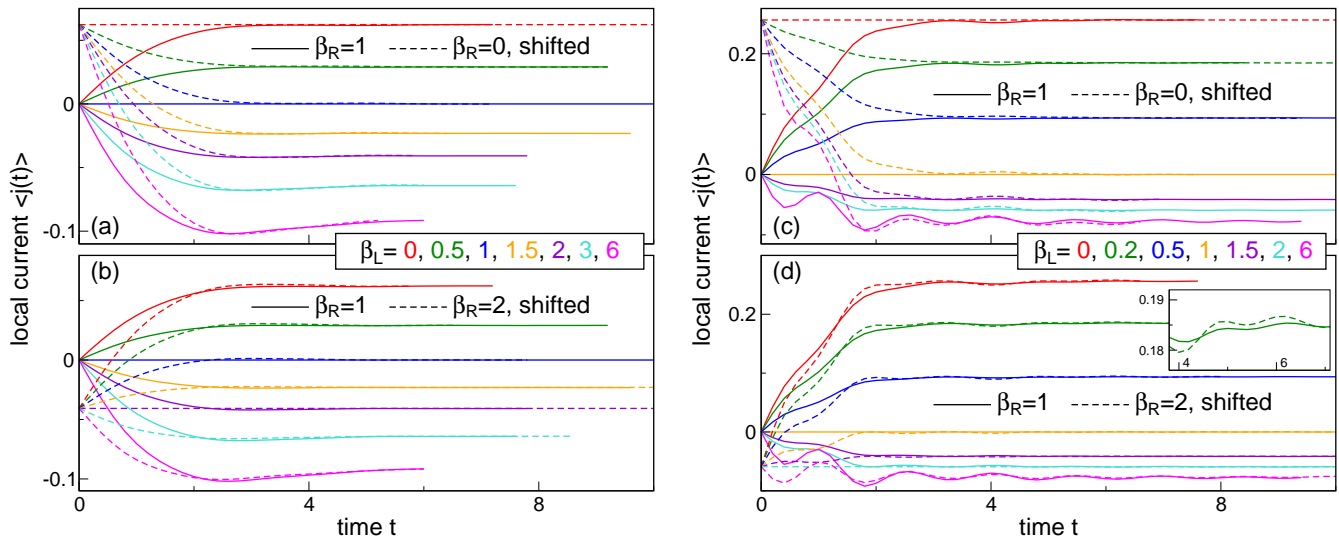

\includegraphics[width=0.48\linewidth,clip]{hubU2.eps}\hspace*{0.02\linewidth}
\includegraphics[width=0.48\linewidth,clip]{hubU8.eps}
\caption{(Color online) Thermal current for the integrable model with fixed $\beta_R\in\{0,1,2\}$ and various $\beta_L$ (increasing from top to bottom). The on-site interaction strength is given by $U=1$ in (a,b) and $U=4$ in (c,d) [the charge gap is of size $\sim0.04$ and $\sim1.17$, respectively]. The inset to (d) shows the data for $\beta_L=0.2$ on a magnified scale. If the curves for the different $\beta_R$ are shifted vertically (by the same amount for all $\beta_L$), they seem to converge to the same long-time asymptote, which thus depends only on $\beta_L$. This illustrates that on the time scales accessible by the DMRG, the stationary state is effectively described by the additivity relation of Eq.~(\ref{ffunc}); violations are smaller than the resolution of the method.}
\label{fhub}
\end{figure*}

From a numerical perspective, it is advantageous not to work with the product state of Eq.~(\ref{state1}) but to model the `bond' between the left and right systems smoothly. This can be achieved by using
\begin{equation}\label{state2}
 \rho_0 \sim e^{-\tilde H}\,,
\end{equation}
and, for the temperature gradient setup, choosing $\tilde H$ as
\begin{equation}
 \tilde H =
 \begin{cases}
  H/T_L & n\leq 0\\ H/T_R & n >0 \,.           
\end{cases}
\end{equation}
Likewise, one can create a vacuum in the right half by adding a large chemical potential $\mu(\tilde n_{n\uparrow}+\tilde n_{n\downarrow})$ for sites $n>0$. It is important to keep in mind that the modified Hamiltonian $\tilde H$ only governs the preparation of the initial state; the real time evolution is always determined by the original $H$ of Eq.~(\ref{hub}). Since Eq.~(\ref{state1}) and Eq.~(\ref{state2}) differ only locally, one expects that they yield the same stationary state at long times, which we have verified explicitly. However, the latter choice is numerically favorable (the DMRG bond dimension increases more slowly); this is reasonable since the initial state is already `closer' to the stationary one if the left and right systems are not fully disconnected initially. Hence, we exclusively employ Eq.~(\ref{state2}) from now on.

\subsection{Density Matrix Renormalization Group}

The thermal density matrices $e^{-H/T}$ as well as the real time evolution operators $e^{-iHt}$ of one-dimensional systems can be determined by virtue of the time-dependent \cite{tdmrg1,tdmrg2,tdmrg3,tdmrg4,tdmrg5,tdmrg6} density matrix renormalization group method \cite{white1,dmrgrev,dmrgrev2}, which in practice can be set up elegantly using matrix product states \cite{mps1,mps2,mps3,mps4}. Finite temperatures \cite{dmrgT,barthel,verstraete,vidalop,tmrg1,metts,trick2a} are incorporated via a purification $|\Psi\rangle_T$ of $e^{-H/T}$, and the state $|\Psi\rangle_T$ can be obtained from the (known) $|\Psi\rangle_\infty$ using an evolution $e^{-H/2T}$ in $\beta=1/T$. Both $e^{-H/T}$ as well as $e^{-iHt}$ are factorized by a fourth order Trotter-Suzuki decomposition. We keep the discarded weight during each individual `bond update' below a threshold value, which leads to an exponential increase of the bond dimension during the real time evolution. In order to access time scales as large as possible, we employ a finite-temperature disentangler \cite{drudepaper,dmrgpaper2}, which exploits the fact that purification is not unique to slow down the growth of the bond dimension. Our calculations are performed using a system size of the order of $L\sim O(100)$ sites. By comparing to other values of $L$, we have ensured that $L$ is large enough for the results to be effectively in the thermodynamic limit.

\section{Currents induced\\ by a temperature gradient}
\label{sec:T}

In this section, we study the energy currents induced by a temperature gradient (defined by $T_{L}$ and $T_R$) in the Fermi-Hubbard model. We exclusively focus on the case of half filling where a charge gap opens for $U>0$ while spin excitation remain gapless. We always average $\langle j_n\rangle$ over the two sites $n$ closest to the boundary where the initial $T$-gradient is applied; this average is denoted as $\langle j\rangle$ from now on.

Fig.~\ref{fcurrent} shows the time evolution of the currents for fixed $U=4$ where the charge gap is of $O(1)$ \cite{hubbook}. As expected, one observes that $\langle j(t)\rangle$ relaxes to a non-zero steady-state value in the integrable case of $V=0$ but decay to zero for non-vanishing nearest-neighbor interactions. However, the time scale on which this decay sets in becomes large if both $T_L$ and $T_R$ are below the gap.

It is \textit{a priori} unclear whether or not the functional form of the steady-state current given in Eq.~(\ref{ffunc}) provides an effective description of thermal transport for the FHM; the reason for this is two-fold. Firstly, the FHM supports ballistic thermal transport in equilibrium \cite{thermconserved,energyhubbard}, but -- in contrast to the XXZ chain -- does not feature a fully-conserved current, $\sum_n [j_n,H]\neq0$. Secondly, it was argued using field theory \cite{doyon3} that Eq.~(\ref{ffunc}) only holds approximately in the gapped phase of the XXZ chain and that small deviations cannot be resolved on the time scales reached in the numerics of Ref.~\cite{thermalpaper}. The half-filled FHM always features a gap whose size is tuneable by the strength of the on-site interaction; hence, it provides an alternate testing ground for the influence of gaps.

\begin{figure}[t]
\includegraphics[width=0.95\linewidth,clip]{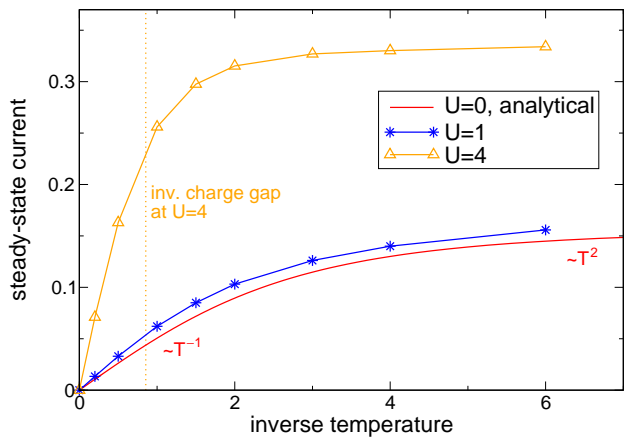}
\caption{(Color online) Temperature-dependence of the function $f(T)$ which governs the steady-state current via Eq.~(\ref{ffunc}). }
\label{fffunc}
\end{figure}

In Fig.~\ref{fhub}, we show the currents flowing in the integrable FHM for different on-site interactions $U=1$ in (a,b) and $U=4$ in (c,d); the size of the charge gap is $\sim0.04$ and $\sim1.17$ in the former and latter case, respectively \cite{hubbook}. Note that at $U=0$, the spin degrees of freedom do not couple, and the problem reduces to that of spinless free fermions where it was shown analytically that Eq.~(\ref{ffunc}) holds. For each $U$, Fig.~\ref{fhub} displays data for three representative temperatures $\beta_R=1/T_R\in\{0,1,2\}$ on the right and varying $\beta_L=0\ldots6$. After shifting the curves for $\beta_R=0$ [panels (a) and (c)] and those for $\beta_R=2$ [panels (b) and (d)] vertically (by the same amount for all $\beta_L$), their long-time asymptotes seem to agree with that of $\beta_R=1$. The same holds true for all other values of $\beta_R=0\ldots6$. This indicates that Eq.~(\ref{ffunc}) provides a good effective description of transport in the FHM; violations cannot be resolved on the finite time scales accessible by the DMRG. A potential reason for this observation might be that Eq.~(\ref{ffunc}) in fact holds exactly in certain regimes (such as large temperatures or small temperature differences). One should note that the parameters of Fig.~\ref{fhub} cover two important limits: The weight of the Drude peak is minimal around $U=1$ \cite{energyhubbard} (the regular contribution to the conductivity vanishes both for $U=0$ and $U\to\infty$), and the temperatures reached at $U=4$ are significantly below the charge gap. No systematically-increasing violation of Eq.~(\ref{ffunc}) can be detected in either case.

\begin{figure}[t]
\includegraphics[width=0.95\linewidth,clip]{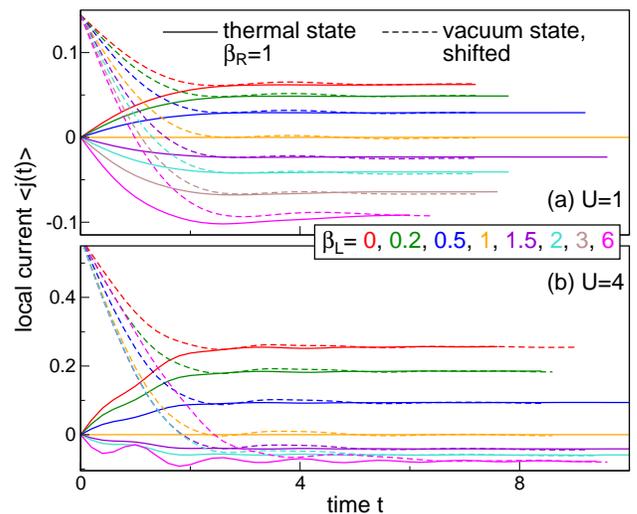}
\caption{(Color online) The same as in Fig.~\ref{fhub}, but comparing the currents flowing into a thermal state at $\beta_R=1$ with an expansion into the vacuum for various $\beta_L$ of the left system. The on-site interaction is (a) $U=1$ and (b) $U=4$. [The curve for $\beta_L=3$ is not shown in (b).]}
\label{fvac}
\end{figure}

In Fig.~\ref{fffunc}, we plot the temperature-dependence of the function $f(T)$ which (to the accuracy of our finite-time numerics) effectively governs the steady-state current. At $U=0$, $f$ is known analytically \cite{free2,bruneau}, and its limiting behavior reads $f(T\to\infty)\sim 1/T$, $f(T\to0)\sim T^2$. In the latter case, the prefactor is determined by the central charge of the corresponding conformal field theory \cite{doyon}, and it was demonstrated explicitly the same holds true for the the XXZ chain in its gapless phase \cite{thermalpaper}. While the large-$T$ limit of the FHM is still described by $f(T)\sim 1/T$, the time scales reached by the DMRG are insufficient to reliably extract the functional form of $f(T)$ for temperatures much smaller than the charge gap; e.g., the data in Fig.~\ref{fffunc} for $U=4$ can be fitted equally (badly) both by a quadratic or an exponential function for $\beta>2$.

\section{Expansion into the vacuum}
\label{sec:vac}

As mentioned above, Eq.~(\ref{ffunc}) can be interpreted intuitively as a form of `effective radiation'. The suggests that the energy currents flowing out of a left system which is prepared in thermal equilibrium at a temperature $T_L$ could approximately be described by $f(T_L)+C$ irrespective of the initial state of the right system as long as the time evolution is governed by an integrable Hamiltonian $H$.

The most important test for this hypothesis is an expansion into a vacuum state (i.e., empty sites), which can be viewed as preparing an equilibrated system in the presence of a sharp potential trap which is then switched off at time $t=0$. In Fig.~\ref{fvac}, we show the currents flowing into the vacuum with the thermal ones for $U\in\{1,4\}$; on the accessible time scales, their steady-state again coincides up to a constant which does not depend on $T_L$. This shows that $\lim_{t\to\infty}\langle j(t)\rangle=f(T_L)+C$ is approximately fulfilled (violations are below the finite-time resolution of the DMRG). Due to the computational complexity of the Hubbard model, we refrain from studying the expansion into various other non-thermal states but will carry out such an analysis for the simpler case of the XXZ chain (see the appendix).

\begin{figure}[t]
\includegraphics[width=0.95\linewidth,clip]{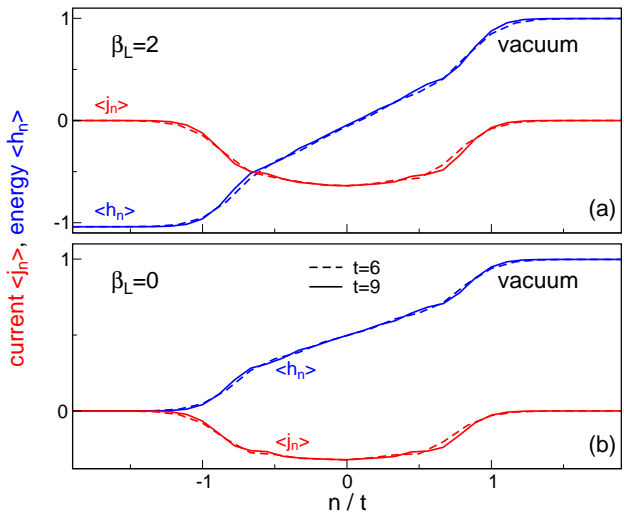}
\caption{(Color online) Full spatial profiles of the energy density $\langle h_n(t)\rangle$ and the energy current $\langle j_n(t)\rangle$ in the FHM at $U=4$ for the expansion of a thermal state with $\beta_L\in\{0,2\}$ into the vacuum for two times $t$.  }
\label{fprof}
\end{figure}

The currents which flow into the vacuum are negative for all $T_L$ (as a reminder, the curves in Fig.~\ref{fvac} are shifted upwards to a constant whose size can be identified by the value at time $t=0$). One can understand this from the fact that formally, the vacuum correspond to a state with a negative temperature, i.e., a state whose energy is higher than the one at $T=\infty$. This is illustrated in Fig.~\ref{fprof} which contains the full spatial profiles of the energy density $\langle h_n(t)\rangle$ as well as the currents $\langle j_n(t)\rangle$ for two different times $t$. Note that after initial transients have died out, both acquire a steady-state form if the position $n$ is rescaled by $t$. Analogous observations were made for the temperature-gradient setup \cite{fagotti}.

\section{Summary \& Outlook}
\label{sec:outlook}

In this paper, we have investigated non-equilibrium thermal transport within the one-dimensional Fermi-Hubbard model. We have shown that the steady-state current flowing out of an initially equilibrated `left' chain can approximately be described via a universal function of its temperature $T_L$, $\lim_{t\to\infty}\langle j(t)\rangle = f(T_L)+C$, irrespective of the initial state on the right. The latter only modifies the constant $C$, where $C=-f(T_R)$ if the system on the right is also thermal with a temperature $T_R$. Violations of this form are below the finite-time resolution of the density matrix renormalization group.

This result is interesting for three reasons: Firstly, it implies that the currents even far out of equilibrium are effectively determined by the linear-response conductance $\partial_Tf(T)$ \cite{doyon}. Secondly, it establishes upper bounds that any analytical solution can be tested against and provides a starting point for the design of phenomenological transport theories \cite{doyon3,rossini,fagotti,doyon6,zotos}. Thirdly, once it becomes feasible to measure energy densities in fermionic quantum-gas microscopes, it should be possible to test this `effective theory' experimentally by preparing a system in equilibrium in the presence of a potential trap, which is then switched off (expansion into the vacuum). 

While the Hubbard model does not feature a fully-conserved energy current, most of the spectral weight of the equilibrium conductivity is concentrated in the Drude peak \cite{thermalpaper}. It would be interesting to generalize the present study to models for which this is not the case.

\begin{figure}[t]
\includegraphics[width=0.95\linewidth,clip]{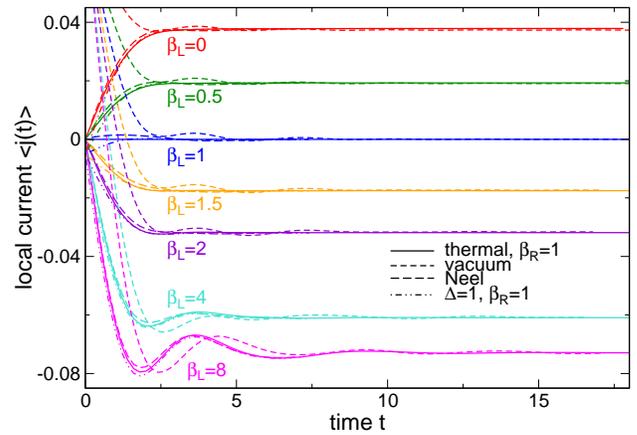}
\caption{(Color online) Thermal current flowing in a XXZ spin chain with $\Delta=0.5$ (this value governs the time evolution) if a thermal state with an inverse temperature $\beta_L$ on the left is connected to different states on the right: (i) a thermal state of the same chain, (ii) the vacuum, (iii) a half-polarized Neel state, and (iv) a thermal state of a XXZ chain with a different $\Delta=1$. The curves for (ii-iv) are shifted vertically (by the same amount for all $\beta_L$). Note that data for (iii) and (iv) could only be calculated up to $t\approx8$ due to the fast growth of the bond dimension for these global quenches.}
\label{fxxz}
\end{figure}

\emph{Acknowledgments} --- Support by the Emmy Noether program of the Deutsche Forschungsgemeinschaft (KA 3360/2-1) is acknowledged.

\section{Appendix: XXZ chain,\\expansion into different states}
\label{sec:app}

We now revisit the non-equilibrium dynamics of a XXZ spin chain whose Hamiltonian reads
\begin{equation}\label{xxz}
H = \sum_{n} \left[\frac{1}{2}\left(S^+_nS^-_{n+1} + S^-_nS^+_{n+1}\right) + \Delta S^z_nS^z_{n+1} \right]\,,
\end{equation}
where $S^\pm=S^x\pm iS^y$, and $S^{x,y,z}$ are spin-$1/2$ operators. Since this model is computationally cheaper than the FHM, one can tackle a larger class of initial states; thermal gradients were already studied in Ref.~\cite{thermalpaper}.

\begin{figure}[t]
\includegraphics[width=0.95\linewidth,clip]{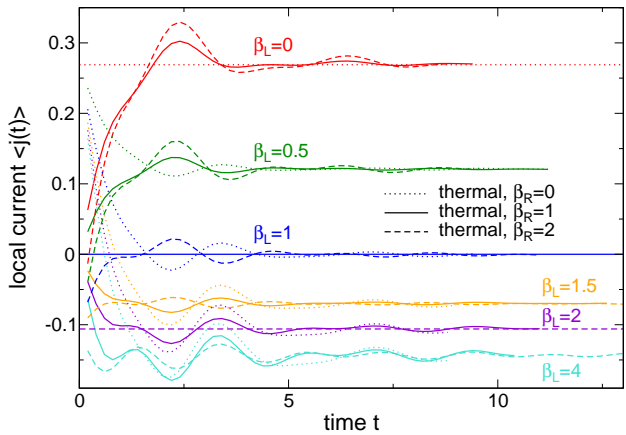}
\caption{(Color online) The same as in Fig.~\ref{fxxz} but in the gapped phase with $\Delta=3$ and for various thermal states on the right. The curves for $\beta_R=0$ and $\beta_R=2$ were shifted vertically.}
\label{fxxz2}
\end{figure}

In Fig.~\ref{fxxz}, we show the currents flowing from a thermal state of a chain with $\Delta=0.5$ and various $T_L$ on the left into several different states on the right: (a) a thermal state, (b) the vacuum (i.e., all spins pointing up), (c) a partial Neel state with a staggered magnetization of $\pm 0.27$ (which we induce by applying a staggered magnetic field of strength $1$ at a temperature $T_R=1$), and (d) a thermal state of a different chain with $\Delta=1$. As a reminder, we note that the time evolution is always governed by the integrable Hamiltonian of Eq.~(\ref{xxz}). The setups (c) and (d) correspond to global quenches, which are numerically challenging; only time scales of $t\approx8$ are accessible in this case. In Fig.~\ref{fxxz2}, we present additional data for the temperature gradient setup in the gapped phase with $\Delta=3$. One observes that the curves seem to approach the same steady-state value if shifted vertically (by the same amount for all $T_L$). This provides further evidence that the asymptotic currents are well approximated by the form $\lim_{t\to\infty}\langle j(t)\rangle = f(T_L)+C$.

\end{document}